\documentclass[prc]{revtex4} % PAPIERFORMAT, STANDARDSCHRIFTGRÖSSE, DOKUMENTENART

\usepackage{eurosym} % ZUR VERWENDUNG DES EURO- SYMBOLS
\usepackage{amsmath} % FÜR MATHEMATISCHE SYMBOLIK
\usepackage{amssymb} % FÜR MATHEMATISCHE SYMBOLIK
\usepackage{dsfont} % FÜR MATHEMATISCHE SYMBOLIK
\usepackage{enumerate} % FÜR AUFZÄHLUNGEN

\usepackage{graphicx} % ZUM GRAPHIKEN EINBINDEN

\usepackage[mathscr]{eucal} % FÜR MATHEMATISCHE SYMBOLIK
\usepackage{longtable} % ZUM ERSTELLEN VON SEHR LANGEN TABELLEN
\usepackage{multirow}

\newcommand{\dd}{\mathrm{d}}
\newcommand{\Dp}{\partial}
\newcommand{\un}{\infty}

\newcommand{\li}{\left}
\newcommand{\ri}{\right}

\newcommand{\abs}[1]{\li| #1 \ri|}

\newcommand{\cen}[1]{\begin{center} #1 \end{center}}

\newcommand{\dis}{{\rm dis}}
\newcommand{\mev}{{\rm MeV}}
\newcommand{\gev}{{\rm GeV}}

\begin{document}
\title{Holographically emulating sequential versus instantaneous disappearance of vector mesons in a hot environment}

\author{R. Z\"ollner}
\author{B. K\"ampfer}
\affiliation{Helmholtz-Zentrum Dresden-Rossendorf, PF 510119, D-01314 Dresden, Germany \\ and \\ Institut f\"ur Theoretische Physik, TU Dresden, D-01062 Dresden, Germany}

\date{\today}

\begin{abstract}
Descent extensions of the soft-wall model are used to accommodate two variants of Regge trajectories of vector meson excitations. At non-zero temperatures, various options for either sequential or instantaneous disappearance of vector mesons as normalisable modes are found, thus emulating deconfinement at a certain temperature in the order of the (pseudo-) critical temperature of QCD. The crucial role of the blackness function, which steers the thermodynamic properties of the considered system, is highlighted.
\end{abstract}

\pacs{}

\maketitle

\section{Introduction}
\begin{figure}
	 \cen{\includegraphics[scale=.6]{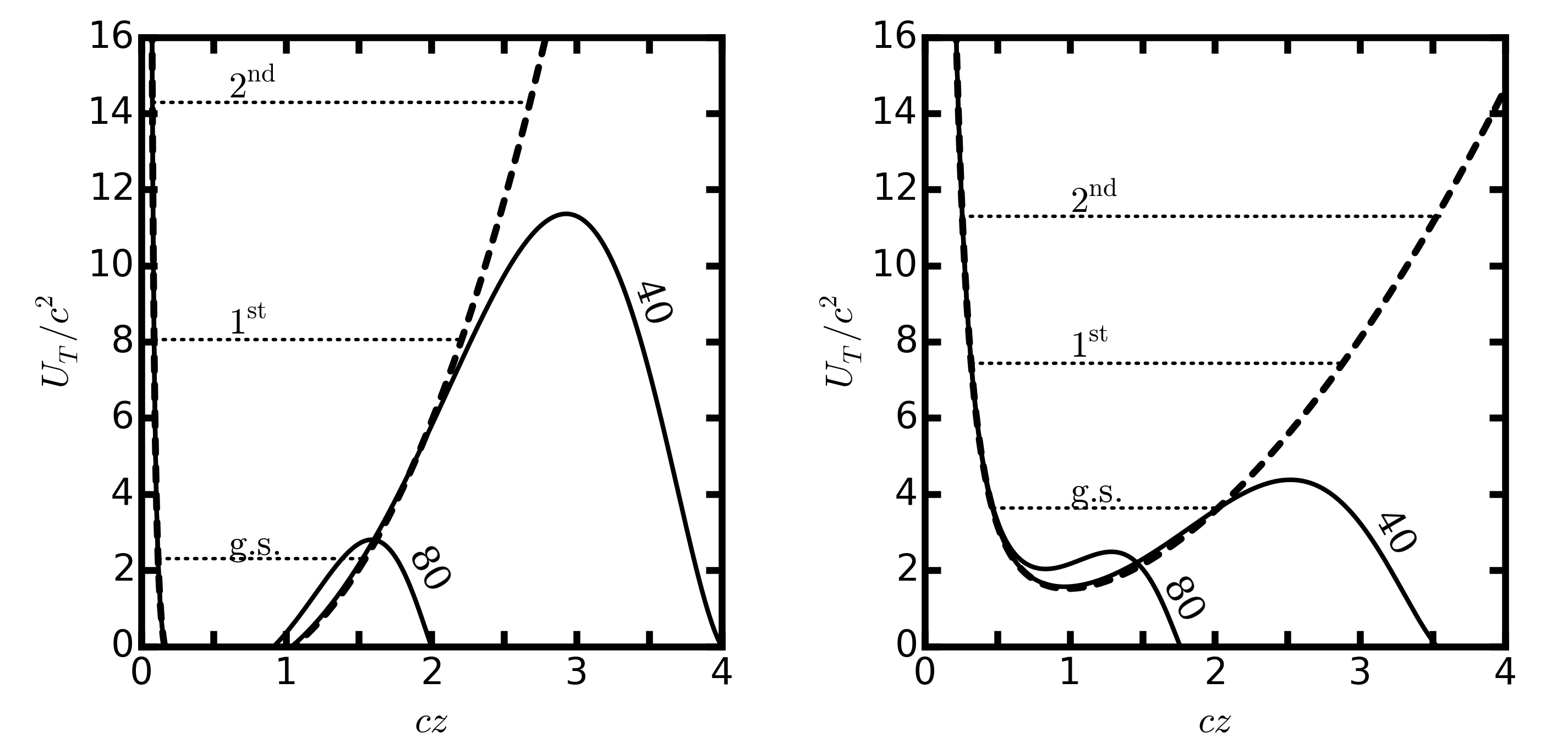}
	 \caption{Scaled Schr\"odinger equivalent potential $U_T/c^2$ (solid curves, labelled by the temperature in units of MeV) as a function of $cz$ (left panel: set 1.2, $cz_H=2$ and 4, right panel: set 2.0, $cz_H=1.75$ and 3.5; the sets are defined in Tab.~\ref{tab1} below). The dashed curve shows $U_0$, i.e. $U_T(z_H\to\un)$ corresponding to $T=0$, with the first three normalisable states (horizontal dotted lines).} \label{abb2}}
	\end{figure}

The quest for in-medium modifications of hadrons initiated a series of investigations, both experimentally and theoretically. The suggestive picture of deconfinement and dissociation of hadrons into quark and gluon constituents was at the origin of relativistic heavy-ion collision experiments seeking and pinning down quark-gluon plasma signatures. The insight that, for 2+1 flavour QCD with physical quark masses, deconfinement happens within a cross over transition at temperatures of $T^{\rm QCD}_c = \mathcal{O}(150 \, \mev)$ \cite{1, 1a} (cf.~\cite{2, 2a} for recent reviews at zero net baryon density) makes such a picture more involved. Also the hints of bound states with charm above $T^{\rm QCD}_c$ \cite{3} point to a complex behaviour of the strongly interacting matter and its constituents or degrees of freedom. \\
At lower temperatures, $T<T^{\rm QCD}_c$, and non-zero baryon density, significant modifications of hadron properties have been expected, too \cite{4, 4a, 4b, 4c, 4d, 4e}. In that regime, the Brown-Rho scaling hypothesis \cite{Brown-Rho1, Brown-Rho2} $m_n \propto \langle\bar q q \rangle ^{x_n}$, where $m_n$ stands for the mass of a hadron and $x_n$ is a specific power of the chiral quark-condensate $\langle \bar q q \rangle$, triggered dedicated studies, most notably turning subsequently in QCD sum rule analyses (cf. \cite{Hatsuda:1991ez, Hilger:2010cn, Hohler:2013eba} for $\rho$ mesons, \cite{Thomas:2005dc} for $\omega$ mesons, \cite{Gubler:2015yna} for $\phi$ mesons, \cite{Hilger:2008jg, Suzuki:2012ze, Buchheim:2014rpa} for $D$ mesons, \cite{Thomas:2007gx, Weise:1994bg} for nucleons, for instance) or hadron-based modelling with relation to chiral restoration effects \cite{Rapp:1999ej, Rapp2}. The particular role of vector mesons (V) roots in their direct decay channel $V \to \bar l l$, i.e. dileptons. Dedicated experiments have been performed, e.g. by HADES \cite{had, had2}, HELIOS-3 \cite{hel}, CERES \cite{cer}, NA60 \cite{na6}, PHENIX \cite{phe, phe2}, STAR \cite{sta, sta2}, uncovering heavy-ion collisions from 1 GeV (fixed target) to 200 GeV (collider), and at LHC (previous experiments up to 5 TeV, collider) one is awaiting also dilepton spectra. While the first expectations focused on effective mass shifts of vector mesons caused by the ambient hot and dense medium, it seems nowadays merely to turn to a broadening of the respective spectral functions. Electromagnetic probes, such as the above mentioned dileptons, are penetrating and carry thus primary information through the strongly interacting medium created transiently in the course of heavy-ion collisions. However, other approaches are conceivable too, for instance, the width determination, via transparency ratios thus accessing cold nuclear matter as ambient medium, cf.~\cite{6} for an example. \\
The advent of the AdS/CFT correspondence \cite{Maldacena,Witten,Gubser} provides a new theoretical tool to access phenomena of in-medium modifications of hadrons. Thereby, hadrons are described as normalisable modes of certain wave equations, or as spectral functions, riding on a gravitational background which in turn is coupled to other fields, most importantly often a dilaton. Issues such as deconfinement and chiral symmetry restoration are addressable within these approaches \cite{7, 7a, 7b}. An important starting point is the soft-wall (SW) model \cite{KKSS}, where a real-valued dilaton field $\Phi = c^2z^2$ and a gravitational warp factor $A= -\ln (z^2/L^2)$ act together to deliver an energy spectrum of normalisable hadron modes corresponding to a mass spectrum $m_n^2 = 4c^2 (1+n)$ where $n=0$ refers to the ground state and $n=1,2,3 \ldots$ enumerates the excitation. Here, $z$ is the holographic coordinate of the five-dimensional Anti-de Sitter space (AdS) with metric determined by the infinitesimal line element squared $\dd s^2 = e^A (\dd t^2 -\dd \vec x^{\,2} - \dd z^2)$. The dilaton (with scale parameter $c$) and warp factor (with the AdS parameter $L$) combine to an effective Schr\"odinger equation type potential $U_0 = 3/(4z^2) +c^4z^2$, and the states follow from a Schr\"odinger equation $\li(\Dp_z ^2 -(U_0-m_n)^2 \ri) \psi =0$. The scale $c$ can be fixed, e.g. by the choice of the ground state $\rho$ meson mass, $m_0 = m_{\rho}$. Remarkable is the Regge type behaviour, $m_n^2 = \beta_0 + \beta_1 n$, with $\beta_0=\beta_1=4c^2$ in the original SW model. \\
One could be tempted to study within an extended approach the impact of small / moderate temperatures, similar to QCD sum rules which are often employed in a low-temperature expansion. Accordingly, the AdS is modified by a black brane to become $\dd s^2 = e^A \li(f(z,z_H) \dd t^2 - \dd \vec x^{\,2} - \dd z^2/f(z,z_H) \ri) $, where the Hawking temperature follows as $T(z_H) = - \Dp_z f \mid _{z=z_H} /(4\pi)$ and the entropy density à$\rm \grave{a}$ la Bekenstein-Hawking is $s(z_H) = \exp\{ 3A(z_H)/2\}$; $z_H$ is the horizon position and $z=0$ marks the boundary of the AdS. 
The Schr\"odinger type equation reads then $\li(\Dp_{\xi}^2 -(U_T-m_n^2) \ri) \psi=0 $ with $U_T(z) = U_0 f^2 +\frac14 (\frac12 \Dp_z A-\Dp_z\Phi) \Dp_z f^2$ and $z(\xi)$, where the tortoise coordinate $\xi$ follows from $\dd z = f(z,z_H) \dd \xi$. Figure \ref{abb2} exhibits the resulting  potential $U_T(z)$ for two values of $z_H$ together with $U_0$, which is nothing but $U_T(z_H \to \un)$.

To make all relations explicitly, we adopt here, for the moment being, the pure AdS blackness function $f=1-(z/z_H)^4$, i.e. $T(z_H) = 1/(\pi z_H)$. The impact of a finite temperature, corresponding to $z_H< \un$, consists in cutting off the right hand side branch of the U-shaped potential (see Fig.~\ref{abb2}), with marginal modifications of $U_T$ relative to $U_0$ for $z<z_M$, where $z_M$ is the local maximum of $U_T$. Clearly, only states with $m_n^2 \leq U_T(z_M)$ could be accommodated in $U_T(z,z_H)$, when ignoring for the moment being the slight deformation of $U_T$ vs. $U_0$ and that properly $\xi$ instead of $z$ should be used. Estimating the right branch of the U-shaped potential by $U_0 \approx c^4z^2$, $z_M = \hat \xi z_H$ with $\hat \xi \approx 1/{1.4}$, $U_T(z_M) \approx U_0(z_M)$, the rough minimal requirement $U_T(z_M) \geq m_{\rho}^2$ to accommodate a state with ground state mass of the $\rho$ meson $m_{\rho}$, becomes $T_0 \leq m_{\rho} \eta^2\hat \xi /\pi$, where we put $c=\eta m_{\rho}$, $\eta =1/2$, and use $T=1/(\pi z_H)$ to arrive at $T_0 \leq m_{\rho}/15 = \mathcal{O}(30 \, \mev)$, a number confirmed by numerical evaluations without approximations. Repeating the analysis for excited states, the above formula reads $T_n \leq m_{\rho} \eta^2 /(\pi\sqrt{1+n})$, i.e.~a sequential disappearance of states upon temperature increase: all states $n > (m_{\rho} \eta^2 \hat \xi /\pi T)^2$ disappear for given $T$; increasing $T$ means decreasing $n$. Obviously, the above ground state disappearance temperature of $\mathcal{O}(30\, MeV)$ can hardly be related to deconfinement at the scale of the (pseudo-) critical temperature from QCD, as already noted in \cite{Colangelo} and exercised in \cite{Col09, Col07, Col08, Col12, Col13, sui1, sui2,7b} for other particle species too. \\
One could argue that the reason of the too small value(s) of $T_0$ is anchored in the use of ad hoc ans\"atze for the dilaton profile $\Phi(z)$, the warp factor $A(z)$ and the blackness function $f(z)$. To test the sensitivity of $T_0$ against moderate variations of $\Phi$, $A$ and $f$ we consider here one-parameter extensions, in particular to show that a blackness function with some relation to QCD thermodynamics is needed to cure the problem of $T_0<T^{\rm QCD}_c$. \\
Our paper is organised as follows. Section \ref{s2} recalls a short derivation of the Schr\"odinger type equation of motion of vector meson modes. In section \ref{sec2} we consider one-parameter extensions of the dilaton profile and the warp factor at $T=0$, i.e. $f=1$. We constrain the parameters to retain a Regge type excitation spectrum. The construction of a modified blackness function is presented in section \ref{sec3}, where also the relation to QCD thermodynamics is briefly discussed, in particular w.r.t. phase transitions. The options of sequential disappearance vs. instantaneous disappearance of vector modes or a combination of both ones are considered in section \ref{sec4}, where we analyse the vector meson spectrum for blackness functions mimicking either a first-order phase transition (section V.A) or a featureless thermodynamic behaviour (section V.B). The discussion of our results, a summary and the avenue toward a consistent approach by solving Einstein equations, instead of relying on ad hoc ans\"atze, can be found in section \ref{sec5}.

\section{Setup: Vector mesons}\label{s2}
We consider the commonly used vector meson action (cf.\ \cite{Brodsky} for a recent recollection)
 \begin{equation}
 S_V = -\frac{1}{4k_V} \int  \! \dd z\, \dd^4 x \,   \sqrt{g} e^{-\Phi(z)} F^2  \label{eq.2}
 \end{equation}
($k_V$: is to be chosen to render $S_V$ dimensionless, but irrelevant in our context;
$\Phi$: dilaton field;
$g$: determinant the metric tensor)
over a Riemann space with infinitesimal line element squared
 \begin{equation}
 d s^2 =  e^{A(z)} \left( f(z) d t^2 - d \vec x ^{\, 2} -\frac{1}{f(z)} d z^2 \right),  \label{ds}
 \end{equation}
where $F_{MN} = \partial_M V_N - \partial_N V_M$ (indices $M, N = 0, \ldots, 4$) is the field strength of a $U(1)$ vector field with the components $V_M$ thought to be dual to the current $\bar q \gamma_{\mu} q$ of the boundary theory; the components of the metric can be read off of (\ref{ds}). 
The warp factor $A$ depends, as the blackness function $f$, only on the holographic coordinate $z$. A simple zero of $f(z,z_H)$ at the horizon, $z=z_H$, ensures the applicability of standard black hole (brane) thermodynamics, i.e.\ the temperature follows as mentioned above and is assigned to the temperature of the boundary system at $z \to 0$ (UV), where $f \to 1$.  \\
In radiation gauge, $V_z = 0$, and Lorenz gauge, $\partial^\mu V_\mu = 0$, the ansatz $V_\mu = \epsilon_\mu \varphi(z) \exp( i p_{\nu}  x^{\nu} )$ with $p_M p^M = m^2$ and $\epsilon_\mu$ as four-polarization vector (indices $\mu, \, \nu = 0, \ldots, 3$) yields the equation of motion from (\ref{eq.2})

	\begin{equation}
	\Dp_z^2 \varphi  + \left(\frac{1}{2}\Dp_z A-\Dp_z\Phi + \frac{\Dp_z f}{f} \right) \Dp_z\varphi + \frac{m^2}{f^2} \varphi=0 
	\end{equation}
which becomes a one-dimensional Schr\"odinger equation by transforming to $\psi(z) =  \varphi(z) \, \exp\{-\frac{1}{2}(A(z) - \Phi(z))\} $ and employing the tortoise coordinate $\xi$ with $\mathrm{d} \xi =  \mathrm{d}z /f $:
	\begin{equation} 
	\li(\Dp_{\xi}^2 -(U_T-m_n^2) \ri) \psi=0,  \label{schr}
	\end{equation}
where 
	\begin{equation}
	U_T = \li(\frac12 (\frac12 \Dp_{z}^2 A-\Dp_{z}^2 \Phi) +\frac14 (\frac12 \Dp_{z} A-\Dp_{z} \Phi)^2 \ri) f^2+ \frac14 (\frac12 \Dp_{z} A-\Dp_{z} \Phi) \Dp_{z} f^2 \label{hotpot}
	\end{equation}
is understood to depend on the coordinate $z(\xi)$. $m_n^2$ follow from normalisable solutions of (\ref{schr}). The zero-temperature case is recovered by $f \to 1$ and $\xi=z$. For a reasoning about the theoretical foundations, see \cite{KKSS}. \\
In a closed complete setting, which goes far beyond (\ref{eq.2}), the warp function $A(z)$, the blackness function $f(z)$ and the dilaton profile $\Phi(z)$ would follow consistently from solving Einstein's field equations and the equations of motion of the involved fields beyond the metric ones.

\section{One-parameter extensions of the dilaton profile and the warp factor} \label{sec2}
At $T=0$, i.e. $f=1$, a modest extension of the SW model for the dilation  profile $\Phi(z)$ and the warp factor $A(z)$ is
 \begin{eqnarray}
 \Phi(z)&=& (c z)^p, \qquad p>1 \label{zp} \\
 A (z) &=& \ln \left(\frac{L^2}{z^2} + \mu^2\right), \qquad \mu^2 \geq 0. \label{eq.4a} 
 \end{eqnarray}
The Schr\"odinger equivalent potential at zero temperature becomes 
 \begin{equation}
 U_0(z) = \frac{1}{2z^2}\li(\frac{3+p(cz)^p}{1+\mu^2c^2z^2}-\frac{3}{2(1+\mu^2c^2z^2)^2} -p(p-1)(cz)^p + \frac{1}{2}p^2(cz)^{2p} \ri). \label{u0}
 \end{equation}

 \begin{table}
 \begin{tabular}{c @{\qquad} c @{\qquad} c @{\qquad} c @{\qquad} c @{\qquad} c @{\qquad} c @{\qquad} c}
 \# & input & $p$ & $\mu$ & $c$ [GeV] & $\beta_0$ [$\gev^2$] & $\beta_1$ [$\gev^2$]& $\beta_2$ [$\gev^2$] \\
 \hline
 1.1 & \cite{Klempt_Zaitsev} & 2.11 & 50 & 0.530 & 0.60 & 1.33 & 0.02 \\
 1.2 & \cite{Ebert}& 2.28 & 50 & 0.501 & 0.59 & 1.49 & 0.05 \\
 2.0 & \cite{BK}  & 1.99 & 0.5 & 0.443 & 0.71 & 0.75 & 0.001 \\
 \hline
 \end{tabular}
 \caption{Optimised parameters to describe the Regge trajectories of Refs.~\cite{BK, Ebert, Klempt_Zaitsev}.} \label{tab1}
 \end{table}
 
\begin{figure}
 \cen{\includegraphics[scale=.7]{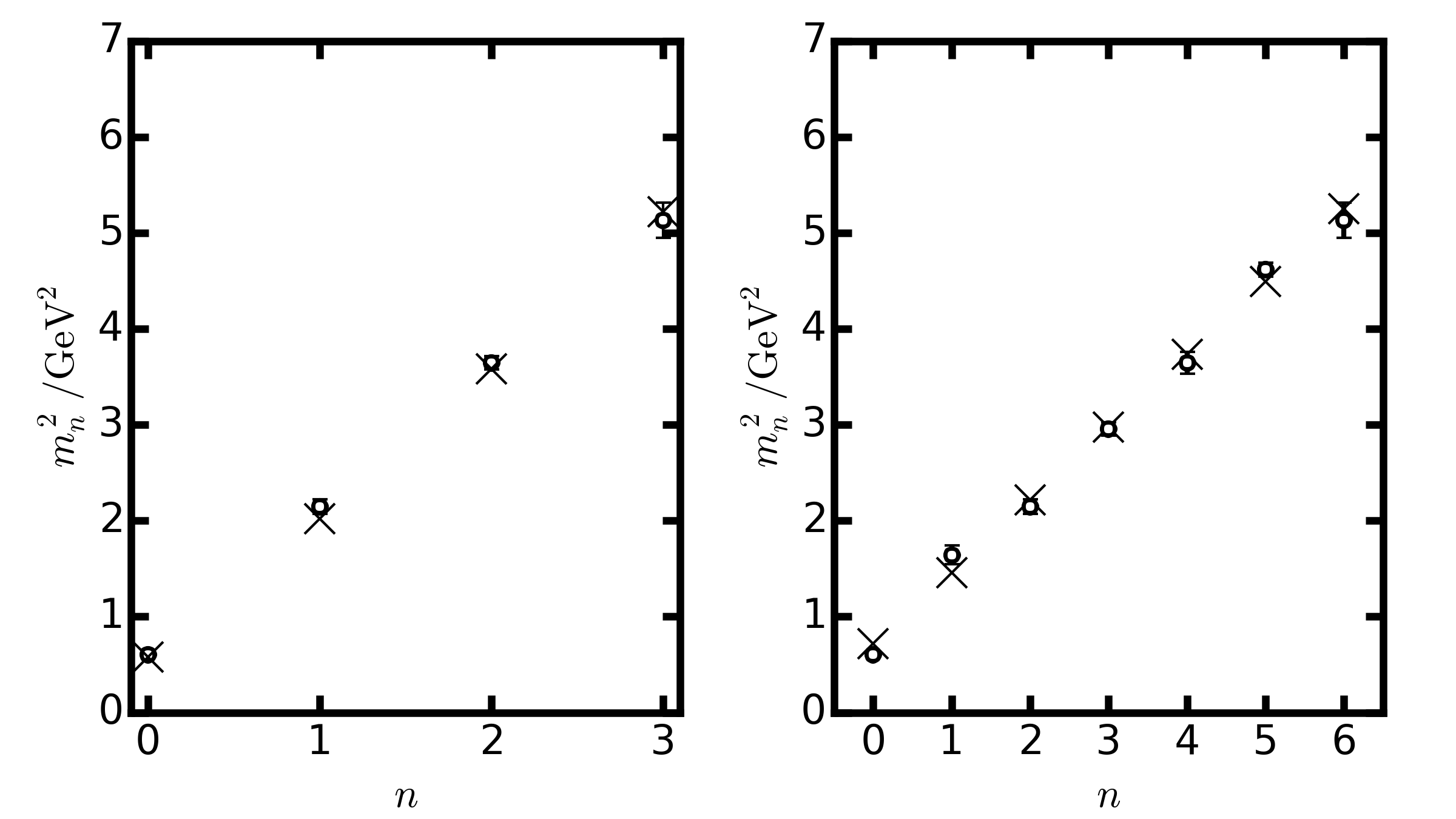}
 \caption{Squared vector meson masses (crosses) from (\ref{schr}) at $T=0$ in comparison with data (circles) for set 1.2 (left panel, \cite{Ebert}) and set 2.0 (right panel, \cite{BK}) from Tab. \ref{tab1}.}  \label{abb1}} 
\end{figure}

We fit the spectra emerging from (\ref{schr}) by
 \begin{equation}
 m_n^2 = \beta_0 + \beta_1 n +\beta_2 n^2 + \ldots 
 \end{equation}
under the requirement of $\abs{\beta_2} \ll \abs{\beta_{0,1}}$ and the smallness of subsequent terms in $\ldots$ which imply a Regge type spectrum. The freedom of $p$, $\mu$ and $c$ (note that we use $L=1/c$ ) can be exploited to get certain wanted spectra. Three possibilities are listed in Table \ref{tab1}. The set 1.1 is optimised w.r.t. the spectrum advocated in \cite{Klempt_Zaitsev}, while 1.2 uses \cite{Ebert} as input. The set 2.0 uses the spectrum in \cite{BK} as input; it is remarkably similar to the SW model. The sets 1.1 and 1.2 are distinguished from 2.0 by a large value of $\mu$ and a slight hardening of the SW dilaton profile. Despite of some difference of the scale parameter $c$, resulting in differences of $\beta_0$, also the Regge slopes $\beta_1$ are fairly different for the different input spectra. Nevertheless, the experimental data are well described, see Fig.~\ref{abb1}.

\begin{figure}
   \cen{\includegraphics[scale=.6]{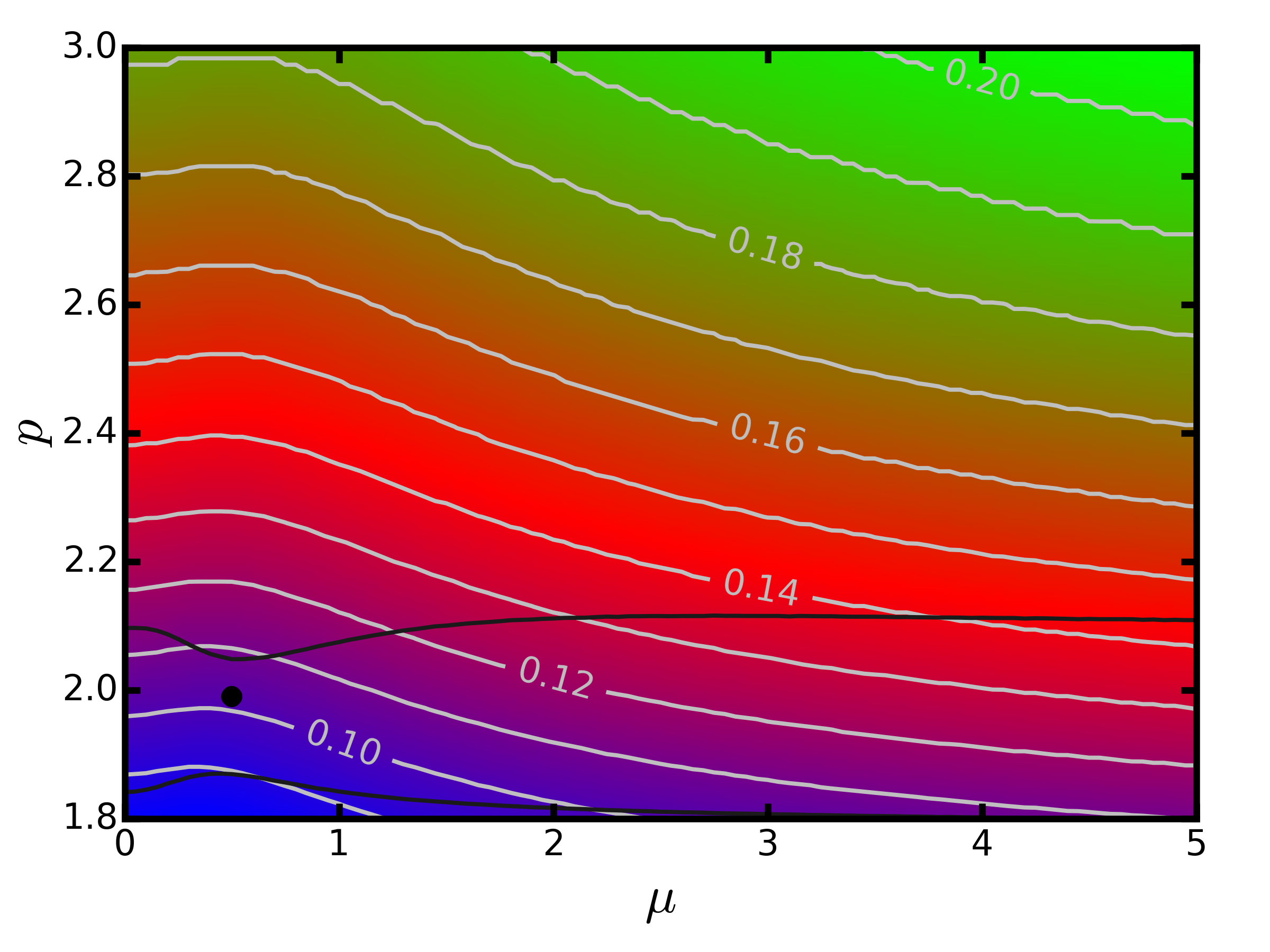}
   \caption{Contour plot (grey curves) of the disappearance temperature $T^{\rm g.s.}_{\dis}/c$ of the ground state over the $p$-$\mu$ plane when employing the potential (\ref{UTpmu}). The black dot at $p=1.99$ and $\mu=0.5$ with $c=443 \, \mev$ belongs to the set 2.0.
   The other optimised parameter points for  sets 1.1 and 1.2 are outside the plot on the $T^{\rm g.s.}_{dis}/c = 0.11$ curve. The black curves limit the corridor wherein $\abs{\beta_2} \leq 0.01 \abs{\beta_1}$.} \label{contour}}
  \end{figure}

The AdS blackness function $f=1-(z/z_H)^4$ gives rise to the potential
	\begin{equation}
	U^{\rm AdS}_T(z) = \li( U_0(z;c,p,\mu)(1-(\pi T z)^4) +2z^2 (\pi T)^4   \li(\frac{1}{1+\mu^2c^2z^2} + p(cz)^p \ri) \ri) (1-(\pi T z)^4) \label{UTpmu}
	\end{equation}
in parametric representation with tortoise coordinate $\xi(z) = \frac14 z_H \ln \frac{z_H+z}{z_H-z} +\frac12 z_H \arctan \frac{z}{z_H}$. We replace here $z_H$ by the temperature according to $z_H = 1/(\pi T)$. \\
Solving the Schr\"odinger equation (\ref{schr}) under the condition to accommodate just the ground state rho with mass of $m_{\rho}=768\,\mev$ yields, for each pair $(p,\mu)$, a disappearance temperature $T^{\rm g.s.}_{\dis}$. That is for $T>T^{\rm g.s.}_{\dis}$ none state is accommodated in the potential. The contour plot of $T^{\rm g.s.}_{\dis}$ over the $p$-$\mu$ plane is exhibited in Fig.~\ref{contour}. Additionally, we have indicated in Fig. \ref{contour} the region wherein $\abs{\beta_2} \leq 0.01\abs{\beta_1}$ (according to the first three states). Only in that region the $T=0$ potential~\eqref{u0} allows for a Regge type excitation spectrum of the lowest few states. For an orientation, also the optimised points $(p,\mu)$ from Tab.~\ref{tab1} are indicated. One observes in fact that these are on contour curves with $T^{\rm g.s.}_{\dis} = \mathcal{O}(50 \, \rm{MeV})$, thus being in the same order of magnitude as anticipated in the Introduction for the SW model. Therefore, the descent modifications (\ref{zp}, \ref{eq.4a}) of the SW dilaton profile and warp factor cannot cure the problem of too low values of $T^{\rm g.s.}_{\dis}$, even over a large range of variations of $p$ and $\mu$. This calls for a proper modification of the blackness function.

\section{Modifying the blackness function} \label{sec3}
Our goal is now to establish suitable modifications of the blackness function in order to evade from the pure scale-free AdS relation $T(z_H) = 1 /(\pi z_H)$. A possible class of blackness functions is provided by

  \begin{equation}
   f(z) = 1-\frac12  \li( \frac{z}{z_H} \ri)^4 -\frac12  \li( \frac{z}{z_H} \ri)^{8(\pi z_HT(z_H)-\frac12)} \label{chagga}
  \end{equation}
 for $T(z_H) \geq \frac1{\pi z_H}$. It fulfils the minimal requirements at boundary, $f(z=0,z_H) = 1$, $(\Dp^i_z f)_{z \to 0} =0 $ for $i=1,2,3$, and has the simple zero at the horizon, $f(z=z_H,z_H)=0$. We consider the following ansatz 

	\begin{equation}
   T(z_H) = \frac1{\pi z_H} + T_{\min} -\frac2{\pi z_{\min}} + \frac{z_H}{\pi z_{\min}^2} \label{Tmin}
  \end{equation}
  
which displays either
	\begin{enumerate}[(i)]
	\item a minimum of $T(z_H = z_{\min})=T_{\min}$ at nearly arbitrarily selectable value $z_{\min}= Z/(\pi T_{\min}) < \un$ (note that (\ref{chagga}) is valid, if $z_H \geq z_{\min}(2-\pi z_{\min}T_{\min})$) with a particular value $Z=2$ yielding the one-parameter form $T(z_H) = 1/(\pi z_H) +  4\pi T_{\min}^2 z_H$ or 
	\item a monotonously decreasing function for $z_{\min} \to \un$ with asymptotic value $T_{\min}$; putting the latter one to zero recovers the pure AdS blackness function.
	\end{enumerate}

The option (i) is of particular interest. Recalling the general relation $v_s^2 = \Dp \ln T / \Dp \ln s$ for the velocity of sound squared one finds $v_s^2 = s(z_H)/T(z_H) \,  \Dp T(z_H) \ \Dp z_H \li(\Dp s(z_H)/\Dp z_H\ri)^{-1}$, i.e. multiplicative factors in both $T$ and $s$ cancel. Suppose a strictly monotonously dropping entropy density $s(z_H)$, then a minimum of $T(z_H)$ at $z_{\min}$ corresponds to a zero of the sound velocity. According to the general discussion in \cite{Kir1,Kir2} this a first-order phase transition. The explicit construction needs the pressure as a function of $T$
 or $z_H$ to find the critical temperature $T_c > T_{\min}$, which we discard here. The important point is that at $T<T_{\min}$ the thermal gas solution without horizon applies and the branch $z_H > z_{\min}$ is unstable. In other words, at least in the temperature interval $0 <T<T_{\min}$ the thermal effects are negligible in leading order of the large-$N_c$ expansion on which the AdS/CFT correspondence is based, i.e.~the $T=0$ solution for the spectrum has to be used. Our expectation is that, at $T_{\min}$ in the order of the QCD transition, the g.s.~and excited states cannot be accommodated in the potential $U_T$, which we refer to as instantaneous disappearance. Such a behaviour emulates deconfinement. In subsection V.A we quantify this issue. \\
Clearly, the ansatz (\ref{Tmin}) does also not allow for an access to the low-temperature effects in the range $0<T<T_{\min}$, but may serve as an indicator of possible effects in the range $T>T_{\min}$ in the case of a given scale $T_{\min}$. Considering $T_{\min}<T_c$ with $T_c$ from QCD, there is nothing in the modelling which refers to a cross over scale $T_c$. \\
Item (ii) also leaves the region $T<T_{\min}$ as thermal gas solution, i.e. in the large $N_c$ expansion, the $T=0$ properties persist in leading order up to $T_{\min}$. \\
The bottom panels of Fig.~\ref{prinzip} exhibit schematically the dependence of the temperature $T$ on the horizon position $z_H$. The regions of the validity of the black brane solution are indicated. In addition, the top panels show the expected behaviour of the states $m_n^2$ as a function of the temperature. The thermal gas solution is in leading order as the $T=0$ solution and does not receive a temperature dependence in contrast to the black brane solution which is affected by the temperature effects (for the sake of simplicity such possible temperature effects are suppressed here). The disappearance points of the states are marked by bullets. For a quantitative account, see the Figs.~\ref{abb5}~and~\ref{abb7} below. \\
We emphasise that beyond the above items (i) and (ii) as some extreme cases further options are conceivable. For instance, $T(z_H)$ may display a stationary or non-stationary inflection point, thus mimicking a second-order or sharp cross-over transition in the ambient thermalised medium \cite{Finazzo1, Finazzo2, Knaute}. Such cases are left for a separate study.

\section{Sequential vs. instantaneous disappearance} \label{sec4}
\begin{figure}
   \cen{\includegraphics[scale=.11]{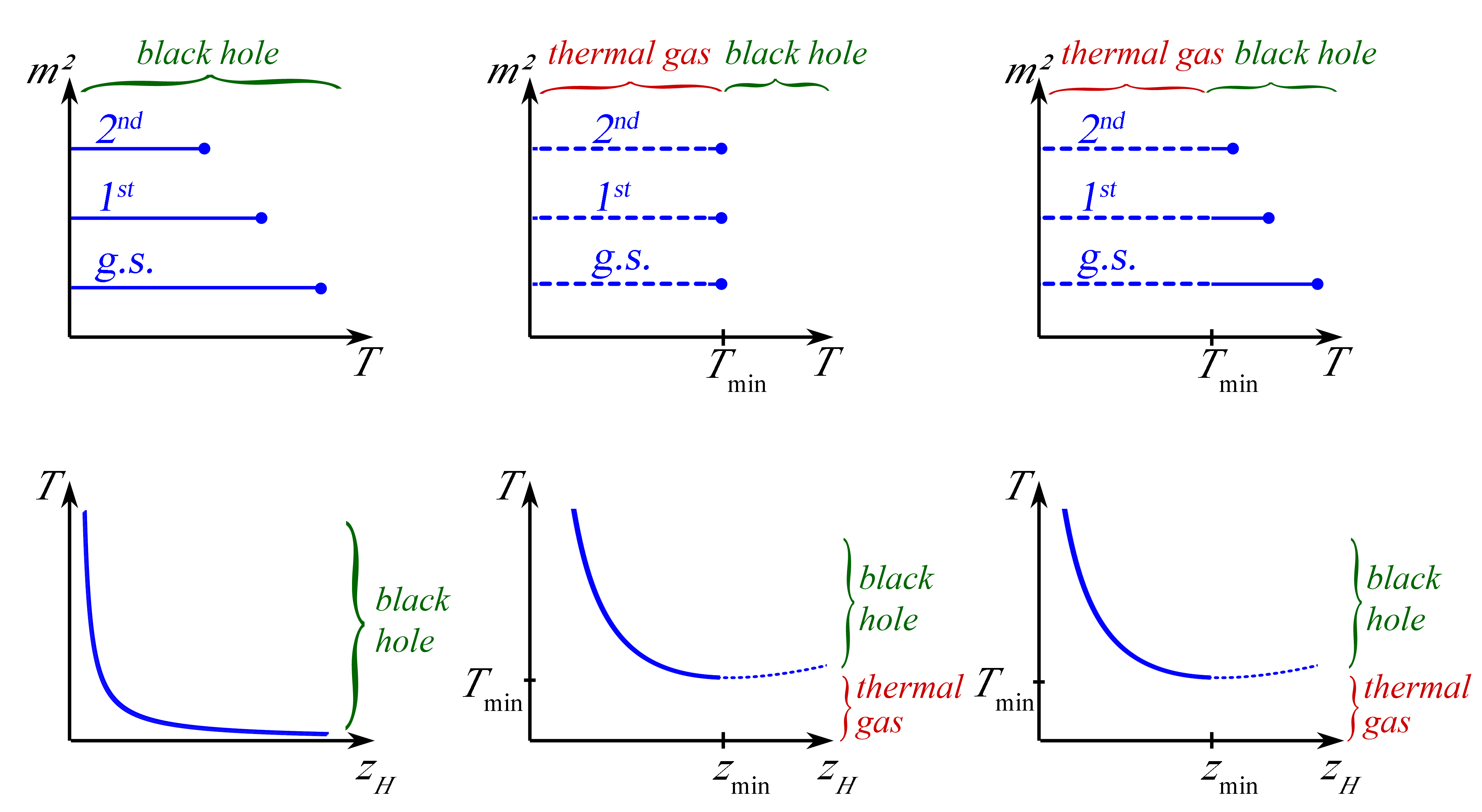}
   \caption{Schematic plot of the temperature $T$ of the ambient medium as a function of the horizon position $z_H$ (bottom panels, unstable sections are dotted) according to (\ref{chagga}, \ref{Tmin}) (left panels: option (ii) for a small value of $T_{\min}$, middle and right panels: option (i)) and the temperature dependence of states (top panels, thermal gas sections are dashed).} \label{prinzip}}
  \end{figure}

We now consider implications of the ans\"atze (\ref{chagga}, \ref{Tmin}) w.r.t.~options for instantaneous and sequential disappearance of modes (cf. top panels of Fig.~\ref{prinzip}) in the Schr\"odinger equivalent potential (\ref{hotpot}) as well as implementing the QCD scale of $T_c$.

\subsection{First-order phase transition: instantaneous disappearance} \label{sec4.1}
\begin{figure}
   \cen{\includegraphics[scale=.69]{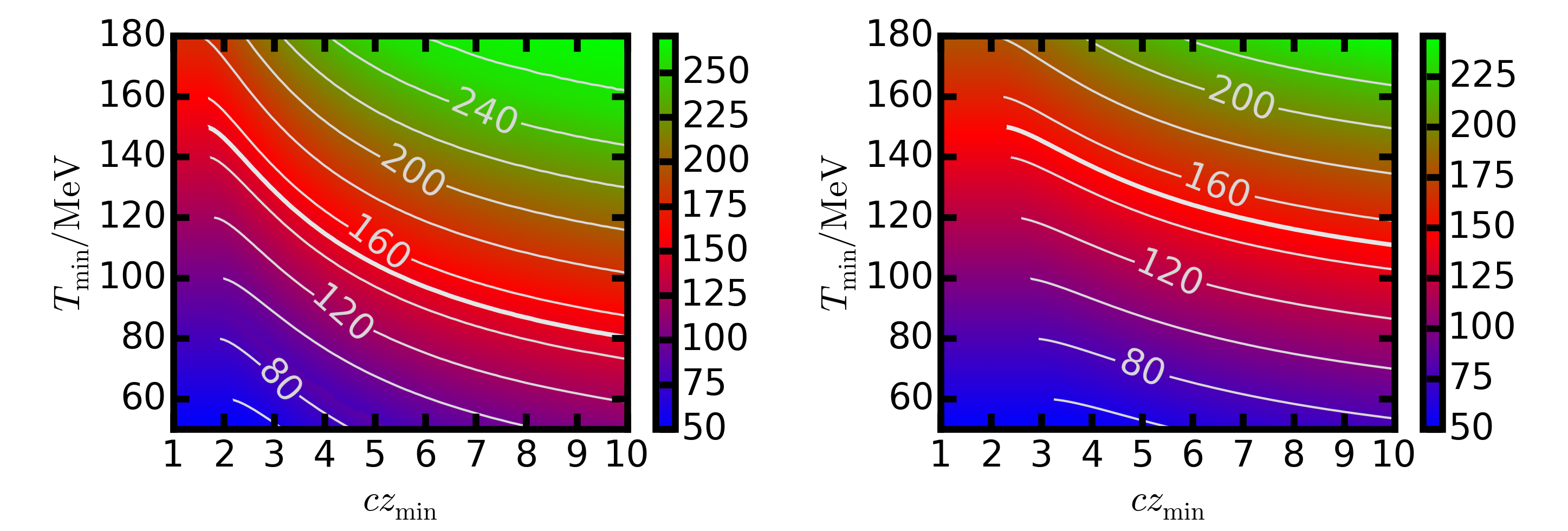}
   \caption{Contour plot of the disappearance temperature $T^{\rm g.s.}_{\dis}$ in units of $\mev$ of the ground state for the parameters of set 1.2 (left panel) and set 2.0 (right panel). The thick grey curve highlights $T_{\dis}^{\rm g.s.} = 150 \, \mev$. If $z_{\min}$ is too small no normalisable states can be found (left part of the panels), i.e.~the contour curves end at $T_{\min}=T_{\dis}^{\rm g.s.}$.} \label{contour2}}
  \end{figure}

Let us consider the temperature $T(z_H)$ according to (\ref{Tmin}) with option (i). In Fig.~\ref{contour2} we exhibit a contour plot of the disappearance temperature $T^{\rm g.s.}_{\dis}$ of the ground state over the $T_{\min}$-$z_{\min}$ plane. That is, for each point $(T_{\min},z_{\min})$ there is a critical value of $z^{\dis}_H$, corresponding to a temperature $T^{\rm g.s.}_{\dis}$ according to (\ref{Tmin}), where the ground state just disappears. Of course, $z^{\dis}_H \leq z_{\min}$ is required to stay on the stable branch. Selecting, for instance, the $T^{\rm g.s.}_{\dis} = 150\, \mev$ curve as QCD relevant, one can infer from Fig.~\ref{contour2} possible combinations of $(T_{\min},z_{\min})$ that deliver the wanted disappearance of all states at $T\geq T^{\rm g.s.}_{\dis}$. In the case of $T^{\rm g.s.}_{\dis} >T_{\min}$, however, it could happen that the potential $U_T(z_H,z_{\min},T_{\min})$ accommodates still a few vector meson states. To ensure the disappearance of all states at $T_{\min}$ one can dial $z_{\min}$ at the endpoint of a given $T^{\rm g.s.}_{\dis}$ contour. Then, in fact, at $T\geq T^{\rm g.s.}_{\dis}=T_{\min}$ no vector meson states are accommodated in the potential $U_T$, and at $T<T^{\rm g.s.}_{\dis}=T_{\min}$ the thermal gas solution, i.e. $T=0$ states, applies. This is the instantaneous disappearance situation, which could be considered as an emulation of a certain deconfinement type. The above mentioned case of $T^{\rm g.s.}_{\dis}>T_{\min} \mathop{=}\limits^{!} T_c$ is related to an emulation, where above the (pseudo-) critical temperature $T_c$ a few lower states could still persist, while higher states already disappeared. Such an option is discussed in the open charm sector \cite{3}; it refers to a partially sequential disappearance. \\
It is worth emphasising that for too small values of $z_{\min}$ no normalisable state can be accommodated in $U_T$; that is the region in Fig.~\ref{contour2}, where no contour curves are displayed. 

\begin{figure}
   \cen{\includegraphics[scale=0.65]{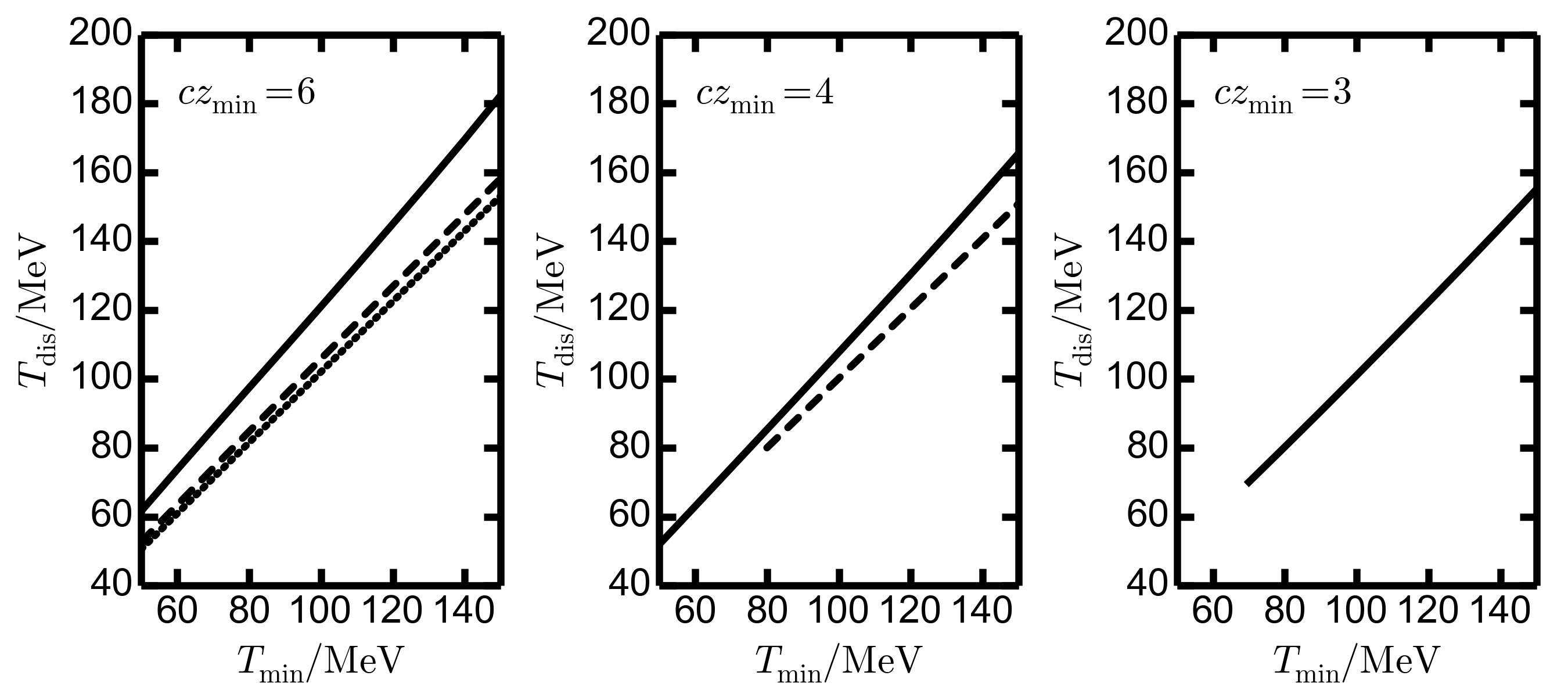} 
   \caption{Disappearance temperature of the first three states (solid lines: g.s., dashed lines: first excited state, dotted lines: second excited state) as a function of the parameter $T_{\min}$ for various values of $z_{\min}$. Left panel: $cz_{\min}=6$ displays again the case of sequential disappearance, i.e.~at given value $T_{\min}$, $T_{\dis}^{\rm g.s.} > T_{\dis}^{\rm 1^{st}} > T_{\dis}^{\rm 2^{nd}}$. 
   Middle panel: $cz_{\min}=4$; up to $T_{\min}=70 \, \mev$, the second and the third mode disappear simultaneously at $T=T_{\min}$, but for greater values of $T_{\min}$ the modes disappear sequentially. 
   Right panel: $cz_{\min}=3$; all states disappear instantaneously up to $T_{\min}=70 \, \mev$; for greater values of $T_{\min}$ the ground state survives.
   The chosen values of $p$, $\mu$ and $c$ belong to set 2.0 of Tab.~\ref{tab1}, i.e. the plots display a cross section through Fig.~\ref{contour2} (right panel) w.r.t.~the ground state.}\label{abb5}}
   \end{figure}

\begin{figure}
   \cen{\includegraphics[scale=.55]{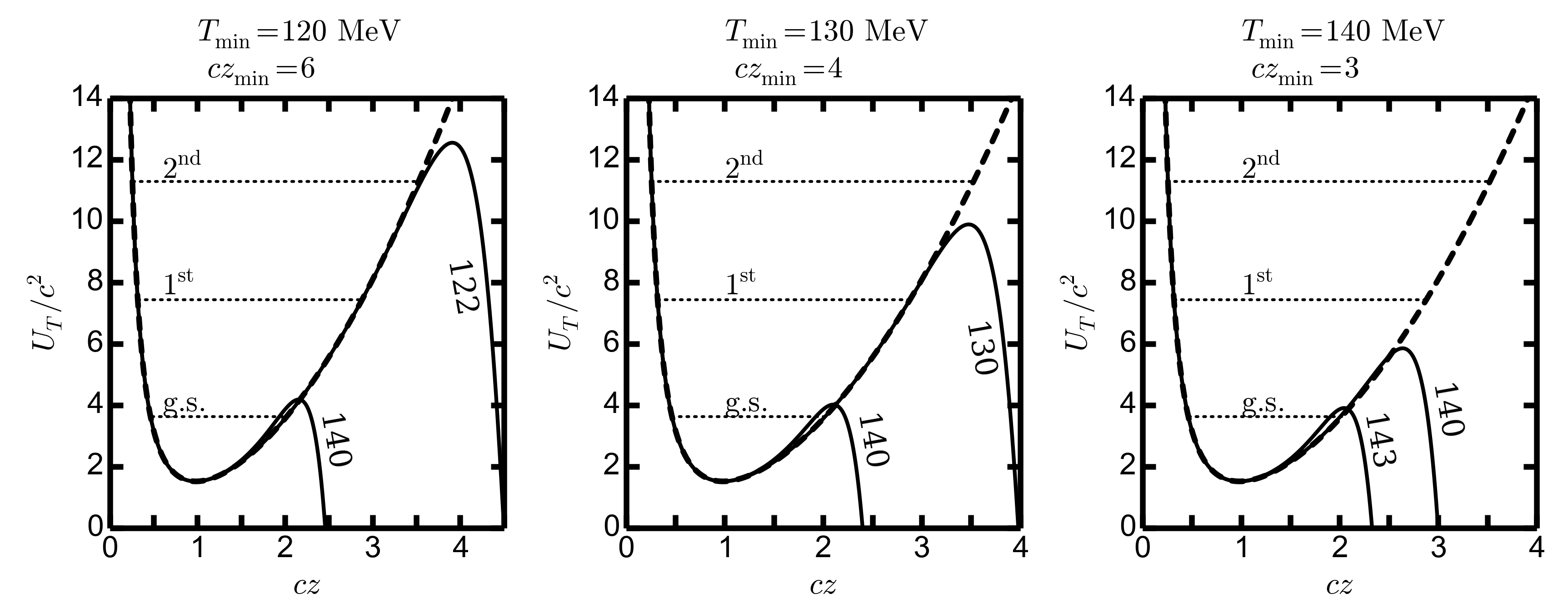}
   \caption{Scaled Schr\"odinger equivalent potentials $U_T/c^2$ (solid curves, labelled by the temperature in units of MeV) and $U_0$ (dashed curves) as a function of $cz$, parameter set 2.0. 
   Left panel: $cz_{\min}=6$, $T_{\min} = 120 \, \mev$, the first three states disappear sequentially.
   Middle panel: $cz_{\min}=4$, $T_{\min} = 130 \, \mev$, the second excited state and all higher ones disappear instantaneously at $T=T_{\min}$ and the remaining two states disappear sequentially.
   Right panel: $cz_{\min}=3$, $T_{\min} = 140 \, \mev$, all excited states disappear instantaneously.} \label{abb3}}
   \end{figure}

Figure~\ref{abb5} exhibits, besides the disappearance temperature $T^{\rm g.s.}_{\dis}$ of the ground state (solid curves), the first (dashed curves) and the second excited states as a function of $T_{\min}$ for three values of $cz_{\min}$. For large values of $cz_{\min}$ (e.g. $cz_{\min}=6$, left panel), all three states persist in the displayed interval of $T_{\min}$, while at smaller values of $cz_{\min}$ (e.g. $cz_{min}=4$, middle panel) the second excited state disappears at all and the first excited state can be accommodated in $U_T$ only for $T_{\min}>80\, \mev$. The first excited state is absent for even smaller values of $cz_{\min}$ (e.g. $cz_{\min}=3$, right panel), and the ground state exists only for $T_{\min}>70\, {\rm MeV}$. At even smaller values of $cz_{\min}$, no normalisable states can be found, as anticipated above. Some examples of Schr\"odinger equivalent potentials are shown in Fig.~\ref{abb3}. In the case of the left panel, $cz_{\min}=6$, $T_{\min} = 120 \, \mev$, the first two excited states disappear sequentially when the temperature rises from 122 MeV to 140 MeV. In the case of middle panel, $cz_{\min}=4$, $T_{\min} = 130 \, \mev$, the third and all higher states have (instantaneously) disappeared already, where the ground state disappears at $T=146 \, \mev$. The right panel, $cz_{\min}=3$, $T_{\min} = 140 \, \mev$, displays the instantaneous disappearance of all excited states, but the ground state.
   
\begin{figure}
   \cen{\includegraphics[scale=0.65]{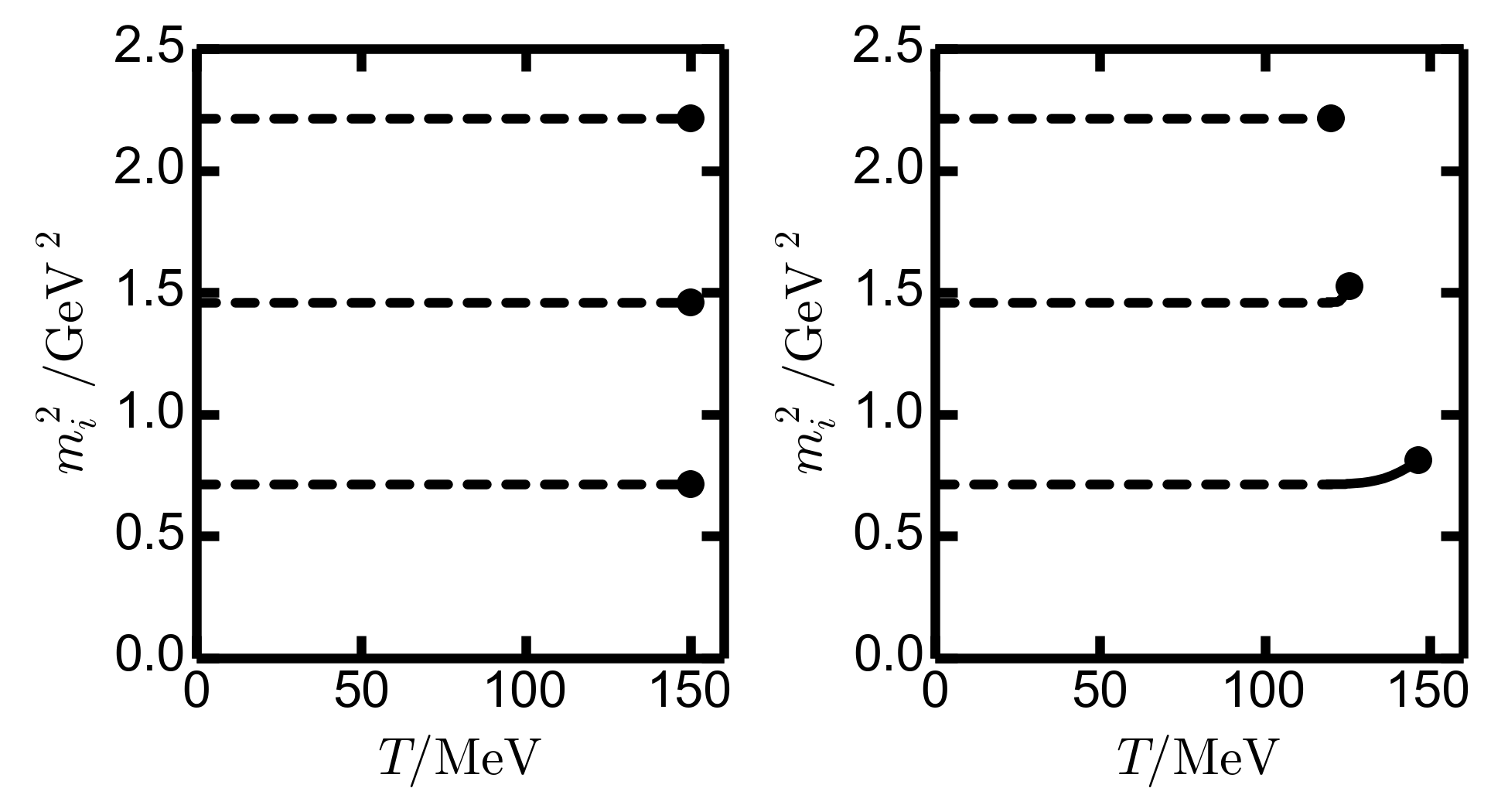}
   \caption{Temperature behaviour of the first three states (set 2.0) for the blackness function $f$ from (\ref{chagga}) with $T$ taken from (\ref{Tmin}) with $T_{\min}=150 \, \mev$ and $cz_{\min}=2$ (left panel) and $T_{\min}=120 \, \mev$ and $cz_{\min}=6$ (right panel). The dashed lines represent the range of $T$ where the thermal gas solution is valid, and the solid curves belong to the black hole solution. This figure provides the quantitative results corresponding to Fig.~\ref{prinzip}, middle and right top panels.} \label{abb7}}
  \end{figure}

These cases can be distinguished by the temperature dependence of the $m_n^2$. The thermal gas leaves the masses of the states independent of $T$ in leading order unless they disappear, see right panel of Fig.~\ref{abb7}. 

\subsection{A featureless example: sequential disappearance} \label{sec4.2}

\begin{figure}
   \cen{\includegraphics[scale=.5]{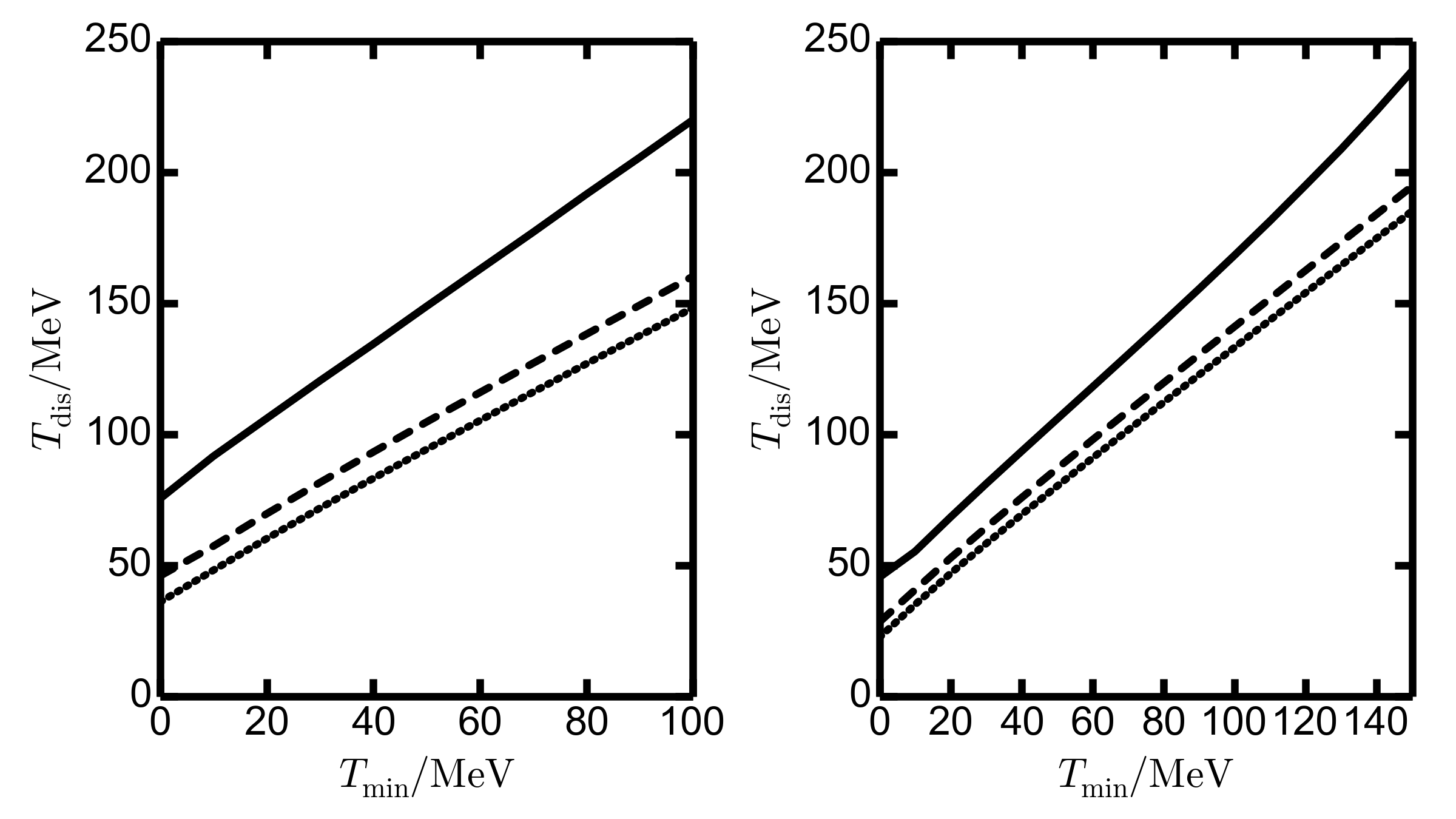} 
   \caption{Disappearance temperature $T_{\dis}$ of the ground state (solid lines), the first excited state (dashed lines) and the second excited state (dotted lines) versus the parameter $T_{\min}$ for set 1.2 (left panel) and  for set 2.0 (right panel) for $z_{\min}\to \un$ in (\ref{chagga}, \ref{Tmin}), i.e.~option (ii).
   Choosing for instance $T_{\min}=55 \, \mev$ (left) or $T_{\min}=\, 95 \, \mev$ (right) the ground state disappears at $T=150 \, \mev$, the first excited state at $T=98\, \mev$ or $T=132 \, \mev$, respectively.} \label{abb4}}
  \end{figure}

We turn now to the option (ii) of the temperature model (\ref{Tmin}) with $z_{\min}\to \un$. Results for the disappearance temperature $T_{\dis}$ as a function of $T_{\min}$ are exhibited in Fig. \ref{abb4}. $T_{\min}=0$ recovers the situation studied in \cite{Colangelo} and discussed in the Introduction, clearly showing the unrealistically small values of the disappearance temperature. Only choosing a sufficiently large value of $T_{\min}$ ensures disappearance temperatures of the order of $T_c$ from QCD. In any case, the sequential disappearance is evident: At given value of $T_{\min}$, the states disappear at $T^{\rm g.s.}_{\dis} > T^{\rm 1^{st}}_{\dis}> T^{\rm 2^{nd}}_{\dis}$. The differences for higher states become gradually smaller pointing to a narrow corridor of temperatures, where they disappear. Such a narrow corridor is in agreement with \cite{Stachel} and seems to be consistent with the successful statistical hadronisation models \cite{Stachel2, CKWX, Becattini}. By some fine tuning one can squeeze the corridor which should contain $T^{\rm QCD}_c$, both for options (i) and (ii).

\section{Summary} \label{sec5}
In summary, we consider in this paper a descent extension of the soft-wall model. Modest generalisations of both the dilaton profile and the warp factor can be adjusted to obtain a rather strict Regge behaviour of radial excitations of vector mesons. Basically, two options for the choice of the intercept and the slope parameter are considered which are distinguished by the selections of states to be attributed to a sequence of radial excitations. When extending this approach to finite temperatures by introducing a black brane, encoded in the blackness function, the pure AdS type form faces the problem of the disappearance of the vector meson states at temperatures significantly below the QCD (pseudo-) critical temperature, as discussed in \cite{Colangelo}. At the origin of such an unwanted effect is the strong deformation of a Schr\"odinger equivalent potential not accommodating normalisable solutions. (Note that the equation of motion of the vector meson modes is a second-order differential equation which can be cast into the form of a one-dimensional Schr\"odinger equation. There is no relation to quark-antiquark bound states since the model operates solely with metric functions, a dilaton field and a $U(1)$ gauge field. Therefore, one cannot speak of the `dissociation' of two-body bound states. Instead we use the term `disappearance' of vector meson states.) \\
To cure that problem we consider an extension of the blackness function keeping the requirements at the AdS boundary. The chosen class of blackness functions allows for several options. One of them (referring to option (ii) in the main text) is the introduction of only one scale, $T_{\min}$, allowing to shift the disappearance temperatures to suitable values, e.g. to $\mathcal{O}(T^{\rm QCD} _c)$. In such a case, for low temperatures, $T<T_{\min}$, the thermal gas solution applies, meaning that the vector meson states are as for $T=0$. For $T>T_{\min}$, the black brane solution applies. Here, the excited states disappear sequentially upon increasing temperature, up to a certain temperature, where also the ground state disappears.\\
Having in mind an emulation of deconfinement we select furthermore a special setting of the blackness function depending on two parameters, $T_{\min}$ and $z_{\min}$, which allow either for an instantaneous disappearance of the vector meson states at $T_{\min}$, to be identified with $T^{\rm QCD}_c$ or another possibility which consists in a parameter choice of $z_{\min}$, where at $T_{\min}$ higher states disappear but a few lower states disappear sequentially upon temperature increase. The ground state disappears at $T^{\rm g.s.}_{\dis}>T_{\min}$. One can dial $T_{\min}=T^{\rm QCD}_c$ or $T^{\rm g.s.}_{\dis}=T^{\rm QCD}_c$ or $T_{\min} <T^{\rm QCD}_c< T^{\rm g.s.}_{\dis}$. (While in the charm, respective charmonium sector the lattice QCD results support the choice $T_{\min}=T^{\rm QCD}_c$, in the light-quark vector meson sector we meet lacking knowledge.) \\
Our very phenomenological study focuses on vector mesons, since these are sources of dileptons which serve as penetrating probes in heavy-ion collisions as important tools to monitor the space-time dynamics. The above described approaches within the extended soft-wall model exhibits only a few options, without stringent contact to the thermodynamics of the ambient medium wherein the vector mesons are immersed. Further options, such as mimicking a cross over or a second-order phase transition are conceivable, but miss a convincing foundation. Instead of ad hoc choices of the dilaton profile, warp factor and blackness function, a systematic approach should be attempted where these functions follow from solutions of Einstein equations and equations of motion. Clearly, in doing so many more hadron species must be included to arrive at a self-consistent description of the thermodynamics of the medium. Moreover, further order parameters beyond the dilaton should be accounted for, most notably the chiral condensate to address properly chiral restoration effects. At $T=0$ such an attempt is reported in \cite{BK}. The extension to $T>0$ looks promising to avoid the ambiguities of our present approach, which stays in the framework of the soft-wall model as a famous application of the AdS/CFT correspondence toward understanding the hadron spectrum and medium modification hereof. The extension to non-zero baryon density is a further important issue.

\begin{acknowledgments}
The work is supported by BMBF and Studienstiftung des deutschen Volkes. We dedicate this paper to the memory of Roman Yaresko.
\end{acknowledgments}

\end{document}